# Reversible and Irreversible Data Hiding Technique

Tanmoy Sarkar[*]
*Corresponding Author*
Neudesic India Pvt. Limited
Hyderabad, India
tanmoy.sarkar@neudesic.com

Sugata Sanyal
Corporate Technology Office
Tata Consultancy Services,
Mumbai, India
sugata.sanyal@tcs.com

*Abstract*— **Steganography (literally meaning covered writing) is the art and science of embedding secret message into seemingly harmless message. Stenography is practice from olden days where in ancient Greece people used wooden blocks to inscribe secret data and cover the date with wax and write normal message on it. Today stenography is used in various field like multimedia, networks, medical, military etc. With increasing technology trends steganography is becoming more and more advanced where people not only interested on hiding messages in multimedia data (cover data) but also at the receiving end they are willing to obtain original cover data without any distortion after extracting secret message.
This paper will discuss few irreversible data hiding techniques and also, some recently proposed reversible data hiding approach using images.**

## I. INTRODUCTION

With the recent advancement of technology people are sharing their private data to each other using internet. The most common way to achieve this is by using encryption i.e. to change the data from one form to another. But the disadvantage of encryption is that it will arouse attention and suspicion among adversaries. Also, in cryptography the key is send as a separate message to receiver so that he/she can decrypt the incoming messages. Such techniques are vulnerable to interception by adversary and subsequent man-in-the-middle attack. In steganography there is no need to send the key as separate message. The secret messages are embedded in innocent message without raising any suspension to adversary.
In this paper we will discuss about few reversible and irreversible data hiding techniques.

### A. Steganography Properties and Implementation steps

The suitability of given steganography system are defined by few following important properties [15]:
1. **Statistical Undetectability**: The stego multimedia message containing covert message should be inconspicuous from adversary
2. **Embedding Capacity**: The maximum amount of data embedded into cover image with minimum cover image distortion (Steganographic capacity).
3. **Minimum False Alarm Rate**: The system reporting the presence of a covert message when in fact no covert message is embedded into the message.

Steganography involves the following steps to embed secret image into cover image:
1. Selection of cover image.
2. Secret message where secret message size is less than cover image size. The larger the difference among two the less the distortion of cover image.
3. Function F (i) used to hide data into cover image.
4. Optional stego key to authenticate the data.

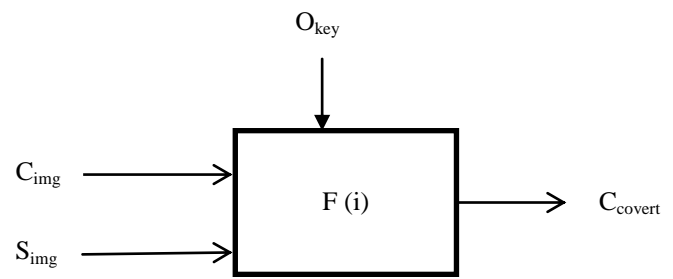

Figure1.1

Where $C_{img}$ is cover media, $S_{img}$ is secret message, $O_{key}$ is optional stego key, and F (i) is steganography function and $C_{covert}$ the stego image.

### B. Digital Watermarking

Similar to Steganography, for authentication of cover message digital watermarking technology is used. The basic difference between Steganography and digital watermarking is that in digital watermarking the covert data is related to cover data but in steganography the covert data is not related to cover data.

### C. Data hiding domains

Spatial Domain Techniques are techniques that operated directly on single pixel of an image.

$$f_i \xrightarrow{Tp(.)} g_i$$

Where $f_i$ is the original image, $g_i$ is the modified image and $T_p(.)$ is the spatial operator defined in a neighborhood p of a given pixel.

Frequency Domain Techniques are operated on frequency of an image.

$$f_i \xrightarrow{f_p} I_i \xrightarrow{-f_p} g_i$$

Where $f_i$ is the original image, $I_i$ is the modified image after applying frequency transformation $f_p$, $g_i$ is the final modified image after implementing inverse transformation $-f_p$.

Wavelet Transform Domain is used in image processing to know the frequency and temporal information at the same time. Using this technique many papers are published which are efficiently store the secret data by shifting histogram and increases data embedded capacity.

## II. STEGANOGRAPHY TECHNIQUES

There is lot of Steganography techniques used to hide data using different cover media like audio, image, video etc. In this paper we will discuss various techniques of data hiding using images and there advantages/disadvantages.

### A. Irreversible data hiding techniques

Irreversible means once the covert image embedded on the original image the original image is lost i.e. from stego image original image cannot be recovered in extraction process.

The simplest method of data hiding in irreversible category is using LSB (Least Significant Bit) insertion. Let us consider the cover message $m_o$ is grey scale message where each pixel is denoted by 8 bits. So, mathematically to replace the first pixel $x_i$ of message $m_o$ with the first bit of cover message $m_c$ is as follows:

$$x_{LSB_i} = x_i \bmod 2$$
$$x_{covert} = x - x_{LSB_i} + m_c \bmod 2$$

The advantage of LSB embedding is its simplicity and the difference is not visible to naked eyes. But this technique has also having lot of disadvantages like LSB encoding is extremely sensitive to any kind of filtering or manipulation of the stego-image. An attack on the stego-image is very likely to destroy the message. An attacker can easily remove the message by removing (zeroing) the entire LSB plane with very little change in the perceptual quality of the modified stego-image. From Fig. 1.2 we can see that after embedding secret message into the cover image there is significant change in original image histogram pattern suggesting it is being distorted.

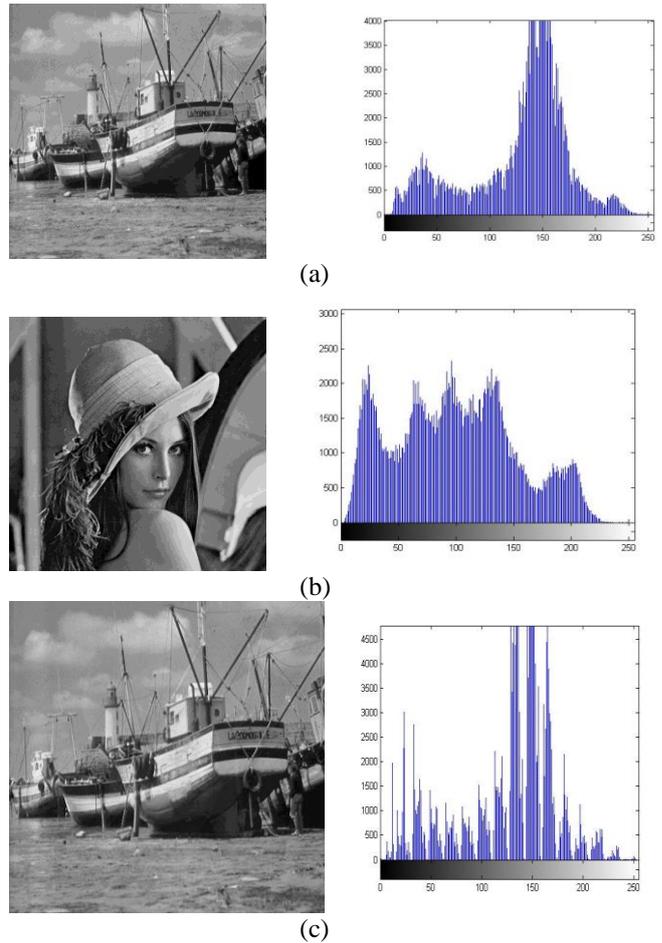

(a)

(b)

(c)

Figure 1.2 LSB Data Hiding Technique (a) Cover Image (b) Image to be embedded (c) Stego Image

Chan et al [1] proposed a novel approach of LSB substitution data hiding in images by using optimal pixel adjustment process. Using this technique the stego image quality is improved than classical approach and also, capacity of embedded secret message increases.

Battisti et al [2] proposed an algorithm based on Fibonacci p-sequence number to increase the bit planes of image and also, reduced the distortion caused by than traditional LSB hiding technique. This paper proposed to decompose the original image into bit planes based on the following Fibonacci p-sequence formula:

$$F(n) = \begin{cases} 0 & n < 0 \\ 1 & n = 0 \\ F_p(n-1) + F_p(n-p-1) & n > 0 \end{cases}$$

After this the covert data is inserted into the selected plane if it satisfies Zeckendrof condition and during extraction it follows the reverse process.

Dey et al [3] [4] proposed an improvement over Fibonacci p-sequence LSB data technique of Battisti et al [2] by decomposing pixel value using two approaches: Prime decomposition and Natural number decomposition technique. Since distortion increases exponentially with the increase of decomposition plane so in these paper the existing number of bit planes increase to a new set of virtual bit planes using decomposition techniques. Unlike Battisti et al [2] these algorithms doesn't use Zeckendrof condition but rather use lexicographical order to define the number system.

Nosrati et al. [6] introduced a method that embeds the secret message in RGB 24 bit color image. This is achieved by applying the concept of the linked list data structures to link the secret messages in the images. First, the secret message that is to be transmitted is embedded in the LSBs of 24 bit RGB color space. The secret message bytes are embedded in the color image erratically and randomly and every message contains a link or a pointer to the address of the next message in the list. Also, a few bytes of the address of the first secret message are used as the stego-key. The disadvantage with this approach is that extra space is required for pointers within image obstructing the possibility of extra message and also, while extraction process if first pointer get corrupted or not traceable on receiver location then whole message becomes unrecoverable.

*B. Reversible data hiding techniques*

Reversible techniques are used to recover the original image from stego image by extracting secret image/message. Areas like medical, military where original image is as important as secret message any vagaries on original image during the transmission can alter the intelligence of original image and affect the overall results.

Ni et al. [5] proposes a novel approach of data hiding using histogram shifting of original image. This algorithm first finds the peak point and zero point in the histogram, records the coordinate of these points and keep the information as overhead information. After this the peak points of the histogram shifted to right by 1 and embed the secret message into the resulting space. Since this approach is reversible the original image can also be extracted by scanning the entire stego image from left to right and apply the process in reverse order. The advantage of this method is that it is simple, reversible and efficient. The disadvantage is that this method will not able to perform when more than one peak and low point exits in the histogram as the overhead message increases. Also, this paper doesn't discuss about the transmission of peak and zero point in histogram to the receiver. If the receiving end doesn't have the knowledge of these points in a histogram then the extraction process fails.

.
Kuo et al. [7] presented a reversible technique that is based on the block division to conceal the data in the image. In this approach the cover image is divided into several equal blocks and then the histogram is generated for each of these blocks. Maximum and minimum points are computed for these histograms and shift the histogram right and left by 1 near the maximum points so that the embedding space can be generated to hide the data at the same time increasing the embedding capacity of the image. A one bit change is used to record the change of the minimum points in location map. In this approach the receiver will extract the location map from the stego image, gets the information about the maximum and minimum points in a histogram and extract the secret message. This method increases the data hiding capacity inside the cover image.

Tian [8] proposes a reversible data hiding technique using difference expansion using LSB. In this method the mean and average value of two neighboring pixel, with small difference value, is calculated. The calculated value is then check to see whether it is satisfying the expandable difference condition (i) and once the condition is passed the new expanded difference is calculated (ii). Finally the secret message is embedded based on the calculated values. This technique also use location map to store the values to know which difference value have been selected which are used to extract the image at the receiving end. This technique significantly improves the capacity of payload message and visual quality of embedded image.

$$2 \times h + b | \leq \min(2(255 - l), 2l + 1) \quad \text{...... (i)}$$
$$h' = 2 \times h + b \quad \text{.................................. (ii)}$$

In the fig. 1.3 we have embedded image (b) into image (a) by using difference expansion. From the histogram of stego image (c) we can see that the secret image is embedded on the difference of near pixel value which are expandable but the pixel having minimum intensity or zero value are not used much in this process.

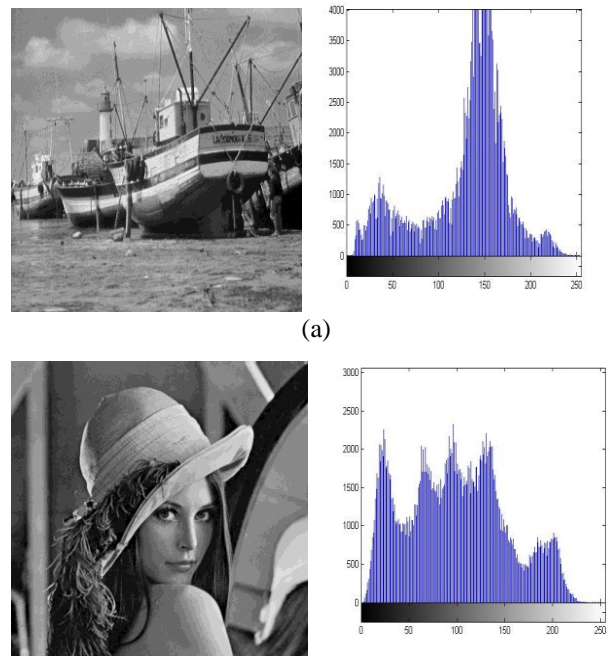

(a)

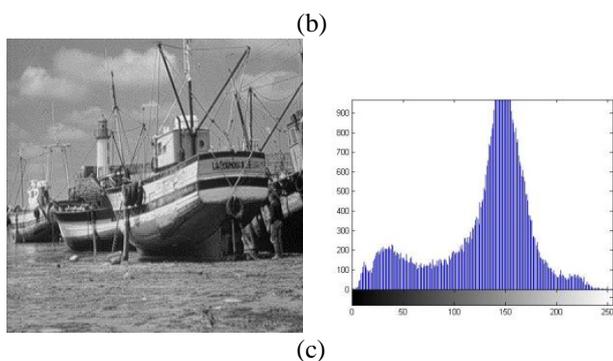
(b)

(c)

Figure 1.3 Difference Expansion using LSB (a) Cover Image (b) Secret Image (c) Stego Image

Yang et al. [9] propose another method of data hiding by embedding a secret message into the low-high (LH) and high-low (HL) sub bands of integer wavelet transform (IWT) domain. This proposed method can effectively embed the covert message without much distortion and also, stego image that get generated equipped with some robustness (protect secret message against image processing operations).The disadvantage with this approach is since this method performed on spatial domain slight changes done on stego image can lead to data extraction failure.

Rosaline et al [13] have put forward a new approach of data hiding by encrypting the original image, compressed the original image using Haar wavelet transform and then embed the secret image into the encrypted compressed data. Using this approach the author provides double protection to secret data.

Xuan et al. [11] [12] proposed reversible data hiding algorithms based on integer wavelet transform (IWT) for high capacity. These algorithms are implemented on IWT domain. The algorithms also, use histogram modification technique, in order to prevent overflow and underflow and embed data into IWT coefficient high frequency sub bands. The first algorithm [12] losslessly compresses some selected middle bit-planes of IWT coefficients in high frequency sub bands to create space to hide data. The second algorithm [13] uses spread spectrum method to hide data in IWT coefficients in high frequency sub bands.

## III. CONCLUSION

In this paper we discussed about various reversible and irreversible steganography techniques and there advantages/disadvantages in spatial and frequency domain. Apart, from these domain some papers has also, been published in software domain [15] where authors try to embed secret message into case-insensitive programming languages, scripts like HTML, Pascal etc.